\documentclass[%
 aip,
 amsmath,amssymb,
reprint, 
]{revtex4-1}

\usepackage{graphicx}
\usepackage{dcolumn}
\usepackage{bm}

\usepackage[utf8]{inputenc}
\usepackage[T1]{fontenc}
\usepackage{mathptmx}
\usepackage{etoolbox}
\usepackage{xcolor}
\usepackage{media9}

\makeatletter
\def\@email#1#2{%
 \endgroup
 \patchcmd{\titleblock@produce}
  {\frontmatter@RRAPformat}
  {\frontmatter@RRAPformat{\produce@RRAP{*#1\href{mailto:#2}{#2}}}\frontmatter@RRAPformat}
  {}{}
}%
\makeatother

\usepackage{hyperref}
\definecolor{linkColor}{rgb}{1,0,0}
\hypersetup{
    pdfborder={0 0 0},
    colorlinks=true,
    urlcolor=linkColor,
    citecolor=linkColor,
    linkcolor=linkColor,
    }
\usepackage{soul}

\emergencystretch=\maxdimen
\begin{document}
\preprint{}

\title[Multislice Ptychography Three-Dimensional Strain Mapping]{Uncovering the Three-Dimensional Structure of Upconverting Core-Shell Nanoparticles with Multislice Electron Ptychography}

\author{Stephanie M. Ribet}
\affiliation{Molecular Foundry, Lawrence Berkeley National Laboratory, Berkeley, CA 94720, USA}
\email{sribet@lbl.gov}

\author{Georgios Varnavides}
\affiliation{Molecular Foundry, Lawrence Berkeley National Laboratory, Berkeley, CA 94720, USA}
\affiliation{Miller Institute for Basic Research in Science, University of California, Berkeley, CA 94720}

\author{Cassio C.S. Pedroso}
\affiliation{Molecular Foundry, Lawrence Berkeley National Laboratory, Berkeley, CA 94720, USA}

\author{Bruce E. Cohen}
\affiliation{Molecular Foundry, Lawrence Berkeley National Laboratory, Berkeley, CA 94720, USA}

\author{Peter Ercius}
\affiliation{Molecular Foundry, Lawrence Berkeley National Laboratory, Berkeley, CA 94720, USA}

\author{Mary C. Scott}
\affiliation{Molecular Foundry, Lawrence Berkeley National Laboratory, Berkeley, CA 94720, USA}
\affiliation{Dept. of Materials Science and Engineering, University of California, Berkeley, CA 94720}

\author{Colin Ophus}
\affiliation{Molecular Foundry, Lawrence Berkeley National Laboratory, Berkeley, CA 94720, USA}
\email{cophus@gmail.com}

\begin{abstract}

In photon upconverting core-shell nanoparticles, structure strongly dictates performance. 
Conventional imaging in scanning transmission electron microscopy has sufficient resolution to probe the atomic structure of these nanoparticles, but contrast, dose, and projection limitations make conventional imaging modes insufficient for fully characterizing these structures.
Phase retrieval methods provide a promising alternative imaging mode, and in particular, multislice electron ptychography can recover depth-dependent information.
Here, we study beam-sensitive photon upconverting core-shell nanoparticles with a multislice ptychography approach using a low electron dose to avoid damage. 
Large strain fields arise in these heterostructures due to the mismatch in lattice parameter between the core and the shell.
We reconstruct both a nanoparticle that appears defect-free and one that has a large break in the side and map the distribution of strain in 3D by computing distortion fields from high-resolution potential images of each slice.
In the defect-free nanoparticle, we observe twisting of the shell, while in the broken nanoparticle we measure the 3D position of the crack, the core, and dislocations. 
These results highlight the advantage of multislice electron ptychography to recover 3D information from a single scan, even under strict electron dose requirements from beam-sensitive samples.

\end{abstract}

\maketitle


Scanning transmission electron microscopy (STEM) imaging is a powerful tool to provide direct characterization of atomic-scale features in materials~\cite{ophus2023quantitative}. 
The small probe size of the converged beam allows for routine imaging of nanoscale defects~\cite{krivanek2010atom, phillips2012atomic}. Conventional bright- and dark-field STEM images are formed by collecting scattered electrons at a fixed angular range with monolithic integrating detectors. 
Despite the great success of these techniques in high-resolution characterization,  conventional imaging modalities possess several limitations.
STEM images are 2D projections of a 3D object, causing overlapping features to potentially be misinterpreted~\cite{midgley2009electron}.
Bright- and dark-field imaging  modalities are relatively dose-inefficient, which limits their applicability for beam-sensitive materials~\cite{chen2020imaging}. 
Additionally, high angle annular dark field (HAADF)-STEM, which is the most common STEM imaging configuration, has a nonlinear dependence on the atomic number of the species~\cite{treacy2011z}. 
This composition-dependence leads to straightforward image interpretation, but dark-field imaging produces little to no contrast for low atomic number elements in most samples. 
For these reasons, there has been growing interest to develop STEM methods which overcome these limitations~\cite{ophus2023quantitative}.

\begin{figure}[htbp]
\includegraphics[width = 0.5\textwidth]{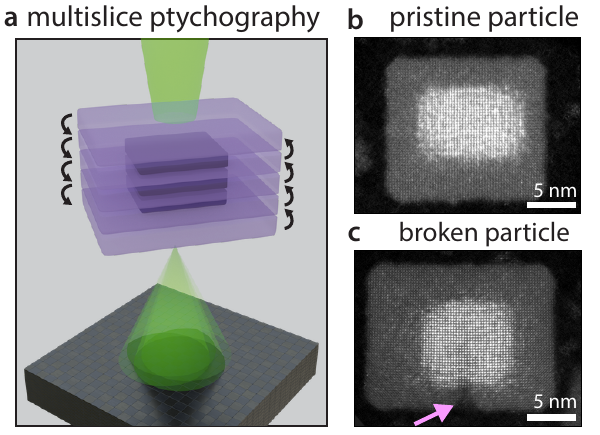}
\caption{\label{fig:intro} (a) Geometry of a multislice ptychographic reconstruction using a defocused electron probe, where the sample is broken into slices along the beam direction. HAADF-STEM imaging of (b) a pristine SrYbF$_5$:1\%Tm@CaF$_2$ nanoparticle, and (c) a particle containing a large crack at the bottom indicated by arrow.}
\end{figure}

\begin{figure*}[htbp]
\includegraphics[width = 0.8\textwidth]{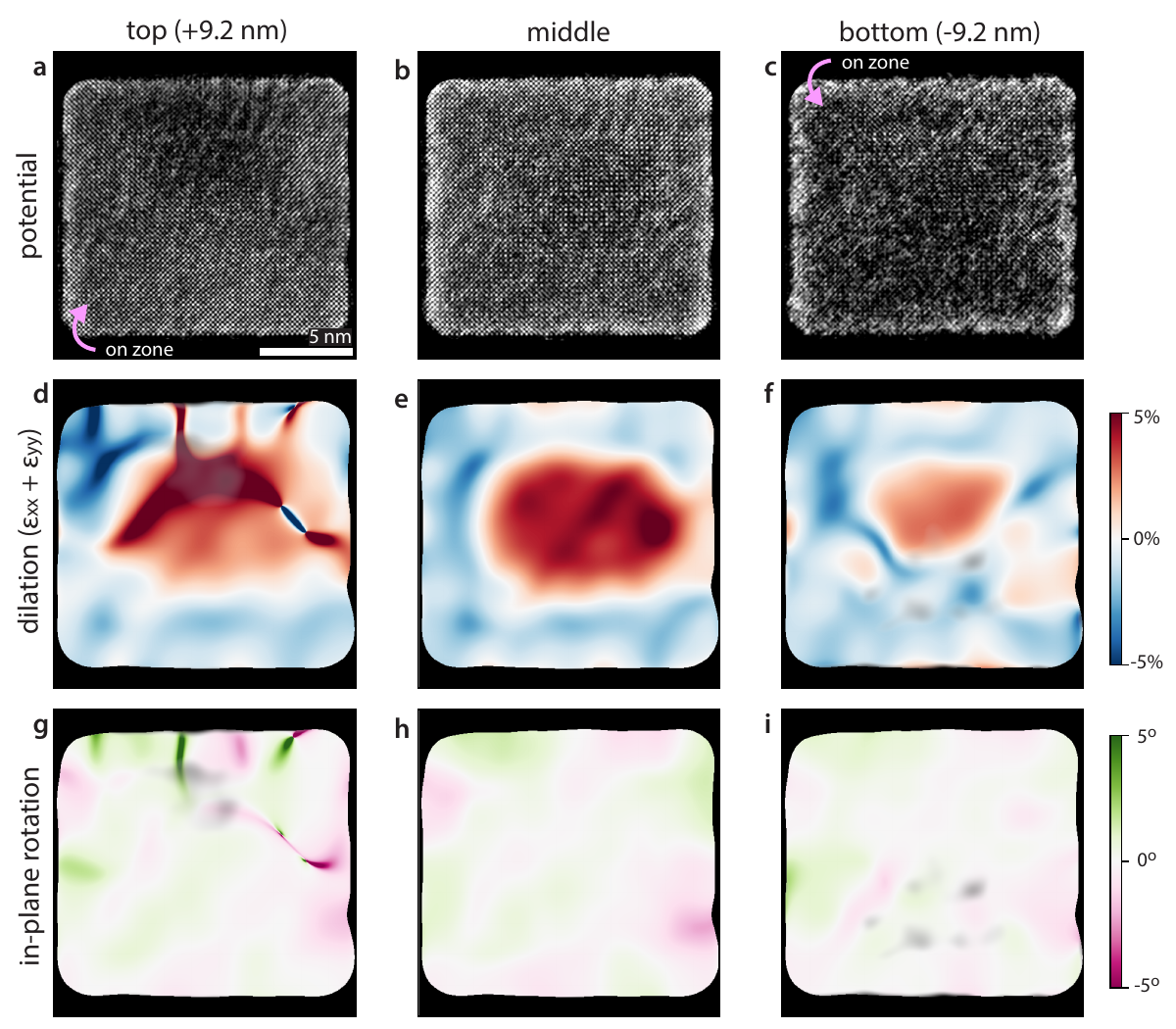}
\caption{
    (a-c) Top, middle, and bottom slice reconstruction of core-shell nanoparticle in Fig.\ref{fig:intro}b, and corresponding (d-f) dilation and (g-i) rotation maps. 
    \label{fig:not_broken}
}
\end{figure*}

Advances in hardware and software have lead to the wide implementation of four dimensional (4D)-STEM techniques~\cite{ophus2019four}. 
Instead of using conventional detectors, 4D-STEM experiments record full diffraction patterns at each probe position, which contain extensive structural information not accessible with conventional imaging modalities. 
Far-field pixelated detectors collect the intensity of the scattered wave, but phase information created by the interaction of the beam and sample is lost. 
Most of the information about the specimen is encoded in the phase of the electron exit wave, especially for weakly scattering samples.
Techniques that recover the phase of the specimen provide dose-efficient characterization of weakly scattering materials, even in a matrix of heavy atoms~\cite{rose1976nonstandard,varnavides2023iterative}.
There are a variety of TEM phase measurement methods including high-resolution transmission electron microscopy, holography, segmented-differential phase contrast, and 4D-STEM based approaches such as differential phase contrast, parallax, and ptychography~\cite{rose1976nonstandard, close2015towards,ophus2016efficient,yang2016enhanced,rodenburg2019ptychography,dunin2019electron, varnavides2023iterative}. 
Iterative electron ptychography algorithms, while more computationally demanding, are particularly powerful phase retrieval techniques.
Ptychographic methods solve both the object and probe with a resolution limit set by the maximum scattering angle, and allow for aberration correction in post-processing and super-resolution imaging~\cite{jiang2018electron, varnavides2023iterative}.

Multislice ptychography is beneficial for strongly scattering samples and allows for recovery of depth-dependent information~\cite{maiden2012ptychographic,chen2021electron}. 
In single slice ptychographic algorithms, the update for each iteration is calculated from a comparison with the experimental diffraction patterns and a projection computed by simple multiplication of the object and probe.
As the object becomes more strongly scattering and the probe travels further through an object, this multiplicative assumption breaks down, leading to artifacts in the reconstruction.
Instead, it is possible to replace the object with thin slices spatially separated along the beam direction (Fig.~\ref{fig:intro}a)~\cite{varnavides2023iterative}. 
In the forward projection, alternating transmission and free-space propagation steps are applied to each slice to account for changes with depth.  
For the backward update, the inverse operations are applied. When properly regularized and converged, multislice ptychography can partition the potential into slices corresponding to 3D information from the object. 
A variety of studies have demonstrated how scanning diffraction data encodes depth information about a specimen, which can be reconstructed with post-processing~\cite{gao2017electron, ophus2019advanced}. 
Recent work has shown how multislice ptychography algorithms can make use of this depth information to characterize 3D information about a sample including defects~\cite{chen2021electron, dong2023visualization, yoon2023, chen2023characterization, gilgenbach2023three}. 
These reconstructions have primarily explored single crystal samples using a high electron dose.

\begin{figure*}[htbp]
\includegraphics[width = 0.8\textwidth]{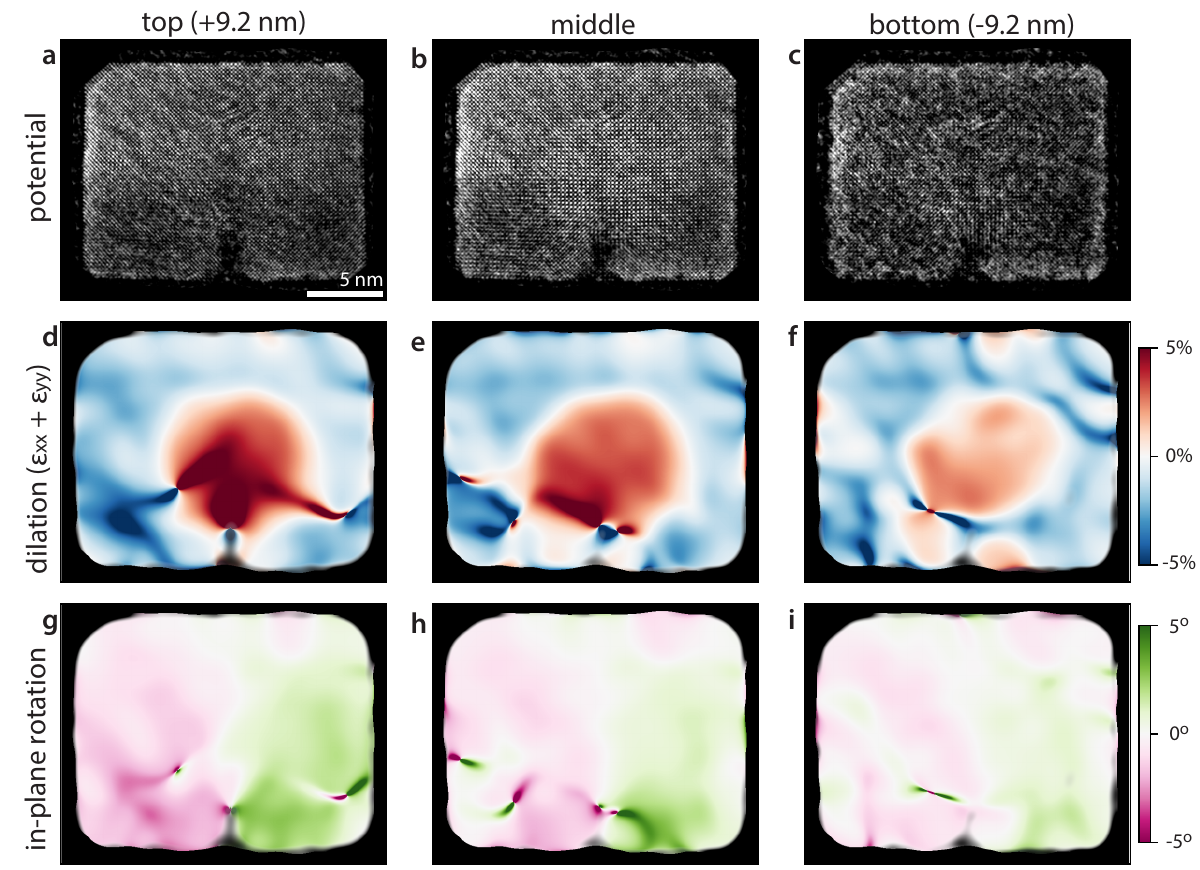}
\caption{
    (a-c) Top, middle, and bottom slice reconstruction of core-shell nanoparticle in Fig.\ref{fig:intro}b, and corresponding (d-f) dilation and (g-i) rotation maps. 
    \label{fig:broken} 
}
\end{figure*}

As a comparison, atomic electron tomography (AET) can provide full atomic resolution in all dimensions~\cite{miao2016atomic}.  
AET experiments use a series of images collected over many tilt projections, but these studies are generally limited to smaller volumes. 
Simultaneous imaging of heavy and light elements in 3D is possible with phase imaging approaches at each tilt~\cite{pelz2023solving}. 
These are technically more difficult experiments, especially under strict dose constraints, and they require much more computational power for reconstruction~\cite{lee2023multislice, varnavides2023iterative}. 

Upconverting nanoparticles are Ln$^{3+}$-doped nanocrystals that convert infrared light to visible and have significant interest for bioimaging~\cite{fischer2020bright, pedroso2021immunotargeting,najafiaghdam2023fully}, optical materials~\cite{mann2021controlled,fernandez2018continuous}, and nanopatterning~\cite{lee2023indefinite}. 
The addition of inert inorganic shells of Ln$^{3+}$-doped cores improves upconversion efficiency by 1-2 order of magnitude~\cite{kwock2021surface, tian2018low}. The dependence of performance on structure underscores the importance of characterizing these nanoparticles at high resolutions.

Core-shell SrYbF$_5$:1\%Tm@CaF$_2$ particles are difficult to characterize with conventional imaging: these beam-sensitive upconverting nanoparticles damage under large doses, and the fluorine atomic constituents are challenging to image.
Thus, we employed multislice ptychography to measure the structure of these core-shell nanoparticles. 
Details about the sample synthesis, experimental measurements, and reconstruction parameters are given in the Supplementary Materials.
We characterized the structure of two nanoparticles, one that appears pristine and one with a visible crack in the side (Figs.~\ref{fig:intro}b-c). 
Because we have reconstructed the atomic potential with depth resolution, we are able to map the depth-dependent strain fields and defect positions.
This work demonstrates the dose-efficient nature of multislice ptychography, and extends this approach to strain and defect mapping of larger, complex material systems.


A HAADF-STEM image of the SrYbF$_5$:1\%Tm@CaF$_2$ core-shell nanoparticle is shown in Fig.~\ref{fig:intro}b, and more images are shown in Fig.~\ref{fig:lowmag} and in prior work~\cite{pattison2023advanced}. 
The compositional contrast of the dark-field image makes it straightforward to distinguish between the core and the shell. 
These nanoparticles have an approximate shell thickness of 3.5 nm, although as show in Fig.~\ref{fig:intro}b the core can be off-center. 
It is not possible to observe the fluorine atoms in the dark-field image.
Most of the particles were intact, but in some cases we observed splitting of the shell, as shown in Fig.~\ref{fig:intro}c and Fig.~\ref{fig:lowmag}b. 
We hypothesize that these cracks form in response to the large strain fields caused by the difference in lattice parameter between the core and shell. 
Both SrYbF$_5$ and CaF$_2$ have a cubic structure. 
SrYbF$_5$ has a unit cell side length of 5.7 \AA{}, while CaF$_2$ has a unit cell dimension of 5.5 \AA{}~\cite{fischer2020bright}. 

We characterized the nanoparticle shown in Fig.~\ref{fig:intro}b using a gradient descent multislice approach implemented in the open-source py4DSTEM toolkit~\cite{varnavides2023iterative,savitzky2021py4dstem}.
We constrain the object to be a positive potential object, and slices from a top, middle, and bottom slice are shown in Fig.~\ref{fig:not_broken}, with all 1.2 nm slices shown in Fig.~\ref{fig:not_broken_slices}. 
Given the low electron dose (5$\times10^3$ e$^-$/\AA$^2$) demanded by these beam-sensitive structures, additional regularization was applied on the probe and object to help with convergence, with details provided in the Supplementary Materials. 
The ptychographic reconstructions reveal the structure more clearly as the flourine atoms become visible.  
Depth dependent information shows that the particle is more complex than previously understood from dark-field imaging.
The contrast in the reconstructed potential is sensitive to tilt. 
Areas where the lattice is aligned with the electron beam, such that the specimen is on zone, appear bright, while tilted regions appear dim. 
The multislice reconstruction uncovers that the tilting of the nanoparticle changes with depth, suggesting twisting, especially of the shell.
In Fig.~\ref{fig:not_broken}a, a slice from top of the particle, the center and bottom portion of the particle are on zone. 
However a slice from further down the nanoparticle (Fig.~\ref{fig:not_broken}c) shows the area in the opposite corner is on zone, as indicated by the arrows.

We have calculated maps of the sample strain using geometric phase analysis of each slice of the reconstruction~\cite{hytch1997geometric}. 
More details about strain and structural analysis can be found in the Supplementary Materials.
Figs.~\ref{fig:not_broken}d-f show the strain dilation (sum of in-plane strain in both directions) from these slices and the corresponding in-plane rotation (Figs.~\ref{fig:not_broken}g-i) from the strain tensor. 
The dominant feature in these images is the large dilation due to the lattice mis-match between the core and the shell.
Additional strain accumulation at the interface between the core and shell and dipoles are present in the top slice indicating more defects near the top.

We apply a similar analysis approach to the broken particle, shown in Fig.~\ref{fig:intro}c, and the results are shown in Fig.~\ref{fig:broken} with all slices shown in Fig.~\ref{fig:broken_slices}. 
In this case, we do not observe as much tilting of the particle, but instead see the evolution of the crack morphology as a function of depth.
In particular, the depth-dependent contrast suggests that the crack is growing towards the top of the particle. 
This feature is also reflected in the rotation maps. 
The magnitude of the rotation in the top slice (Fig.~\ref{fig:broken}g) is much larger than in the middle (Fig.~\ref{fig:broken}h) or bottom (Fig.~\ref{fig:broken}i), suggesting the particle is being pulled apart as the crack expands. 
The dilation maps also show multiple strain dipoles corresponding to lattice dislocations, and more compressive strain near the break.

\begin{figure}
 \includegraphics[width =0.5\textwidth]{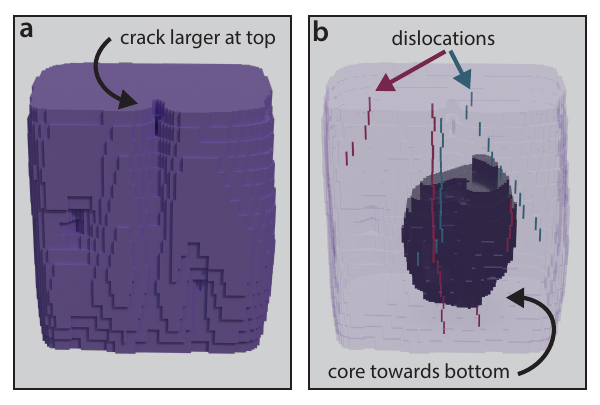}
\caption{(a) Surface of particle in Fig.~\ref{fig:broken} estimated from particle edge in potential slices. 
(b) Internal structure shows shell (purple) and core position (black) and dislocations (in blue and red).
\label{fig:defect} }
\end{figure}

To better understand the architecture of the broken particle, we estimated its 3D structure. 
Fig.~\ref{fig:defect}a shows the surface of the particle, which was calculated using the magnitude of the Fourier peaks from the lattice. 
In agreement with the rotation maps in Fig.~\ref{fig:broken} the crack is largest towards the top of the particle.
We can probe the location of the core relative to the shell by measuring an order parameter corresponding to the relative intensities of the two sublattices with different compositions (measured from the Bragg peak intensities of the image Fourier transforms), which is shown in Fig.~\ref{fig:order}. 
From this analysis, we find the core is also off-center along the beam direction direction, positioned closer to the bottom of the particle. 
Finally, we can also plot the position of dislocations with in-plane Burger vectors in each slice (Fig.~\ref{fig:dislocation}). 
We use Bragg filtering on two of the lattice peaks as indicated in red and blue respectively (Fig.~\ref{fig:fft}) to plot the dislocation positions in each slice (Fig.~\ref{fig:defect}b).
We find dislocations forming towards the edge or interface between the core and shell in the nanoparticle, with some traveling through the full depth.

Work with upconverting nanoparticles which are exceptionally sensitive to changes in quenching~\cite{lee2023indefinite} has shown that defects at the core/shell interface underlie photoswitching~\cite{lee2023indefinite}, and that minor variations in shell thickness can mediate large changes in upconversion efficiency~\cite{kwock2021surface}, although the origins of both effects are poorly understood~\cite{kwock2021surface}.
The methods described here for characterizing the core-shell interface offer a promising route for exploring mechanisms that arise from interfacial strain or structural defects.

This work highlights both the limitations and benefits of multislice ptychography. 
The maximal achievable resolution along the beam direction in multislice ptychography is currently limited to the nanometer scale~\cite{chen2021electron}. 
Features tend to get more blurred out and lengthened along the direction of travel of the electron beam. 
This artifact makes quantification of the absolute aspect ratio and of the termination of top and bottom surfaces challenging without additional information. 
Both a larger convergence angle and a higher electron dose would  better resolve out-of-plane features. 
Collecting additional scans at other tilt angles would be especially valuable to provide information about the changing morphology of the particle along the direction of propagation of the beam.
Because of the limited depth resolution of this study, we can only measure the in-plane components of the strain tensor, though we have demonstrated the depth-dependence of these deformations.
Despite these limitations, this work demonstrates how multislice ptychography can resolve out-of-plane information structural features from a single scan at relatively low dose. 
In particular, we emphasize how multislice ptychography lends itself to the characterization of large, heterostructures and defects in materials.

\section*{Acknowledgements}

CO and SMR acknowledge support from the US Department of Energy Early Career Research Program.
GV acknowledges support from the Miller Institute for Basic Research in Science.
Work at the Molecular Foundry was supported by the Office of Science, Office of Basic Energy Sciences, of the U.S. Department of Energy under contract number DE-AC02-05CH11231. 
This research used resources of the National Energy Research Scientific Computing Center, a DOE Office of Science User Facility supported by the Office of Science of the U.S. Department of Energy under Contract No. DE-AC02-05CH11231.


\section*{Competing Interests}

The authors have no conflicts to disclose.

\newpage
\clearpage

\section*{References}
\bibliography{references}

\newpage
\clearpage

\setcounter{section}{0}
\setcounter{page}{1}
\renewcommand\thepage{S.\arabic{page}} 

\setcounter{figure}{0}  
\renewcommand\thefigure{S.\arabic{figure}}

\section{Methods}
\subsection{Nanoparticle synthesis and sample preparation}

Upconverting core-shell nanoparticle cores (10-nm $\alpha$-phase SrYbF$_5$ doped with 1\% Tm$^{3+}$) were synthesized as described elsewhere~\cite{fischer2020bright}.  
Epitaxial $\alpha$-phase CaF$_2$ shells were overgrown on the SrYbF$_5$ cores following methodology previously described, with minor modifications~\cite{fischer2020bright, pedroso2021immunotargeting}.
Core-shell nanoparticles were drop-cast on a TEM grid with ultra-thin carbon support.

\subsection{Data acquisition}
4D-STEM data was acquired on the TEAM 0.5 double-aberration-corrected microscope at the Molecular Foundry with the 4D Camera running at 87,000 frames per second~\cite{ercius20234d}. 
The microscope was operated at 300kV with a 25 mrad convergence angle and approximately 5$\times10^3$ e$^-$/\AA$^2$. 
Given the large probe size after adding about 30 nm of defocus to the probe, a real space step size of 2\AA{} was used for data acquisition. 
After the 4D-STEM scan, an orthogonal scan pair of in focus dark field images were collected. They were drift corrected in post processing to create high-resolution dark field images~\cite{ophus2016correcting}.

\subsection{Ptychography reconstructions}

\begin{figure*}[h!]
\includegraphics[width=\linewidth]{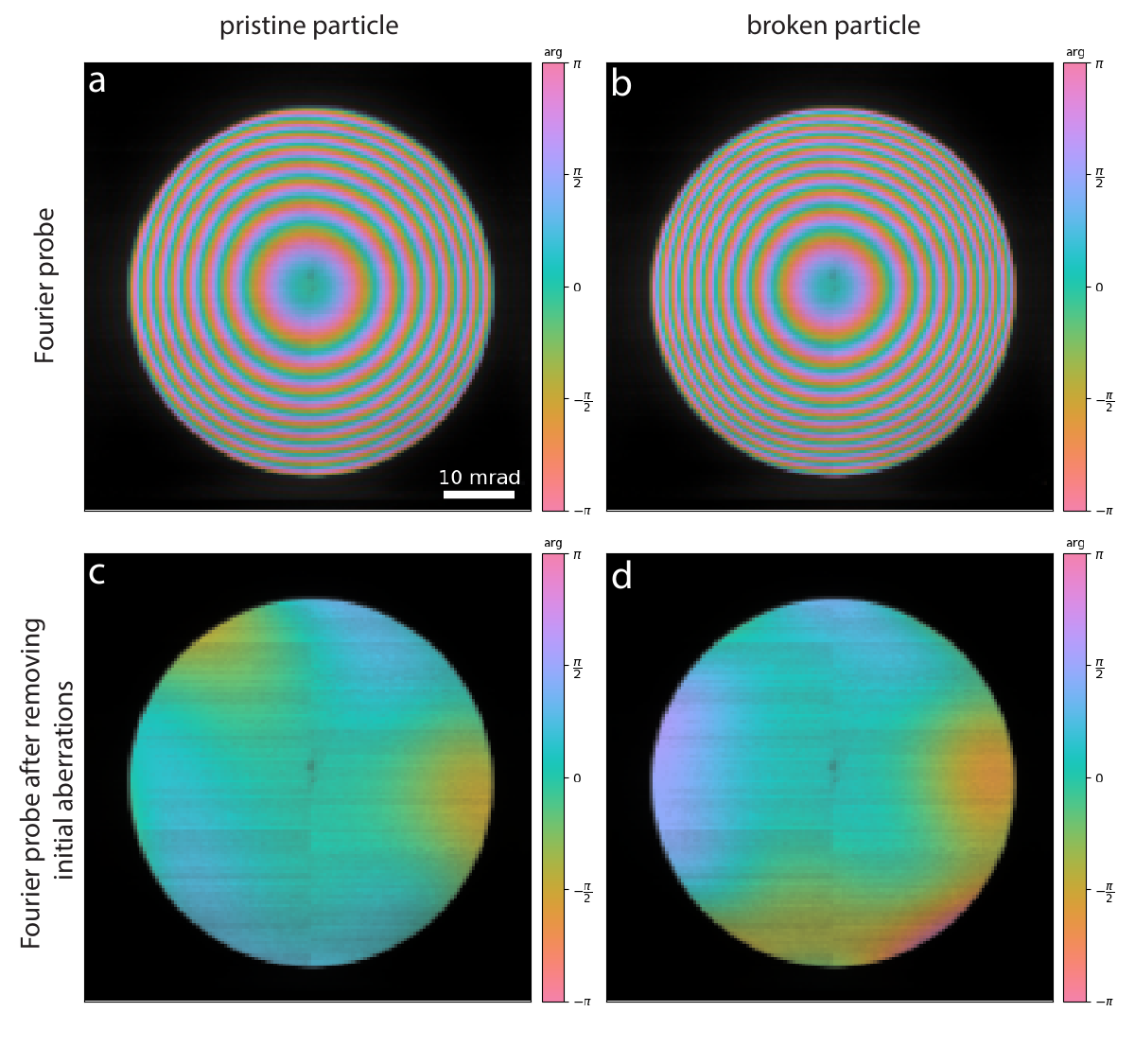}
\caption{(a-b) Reconstructed Fourier probes from Fig.~\ref{fig:not_broken} and Fig.~\ref{fig:broken} respectively. (c-d) Corresponding Fourier probes with initial aberrations removed to highlight changes to probe during reconstruction.}
\label{fig:probes}
\end{figure*}

\begin{figure*}[h!]
\includegraphics[width=\linewidth]{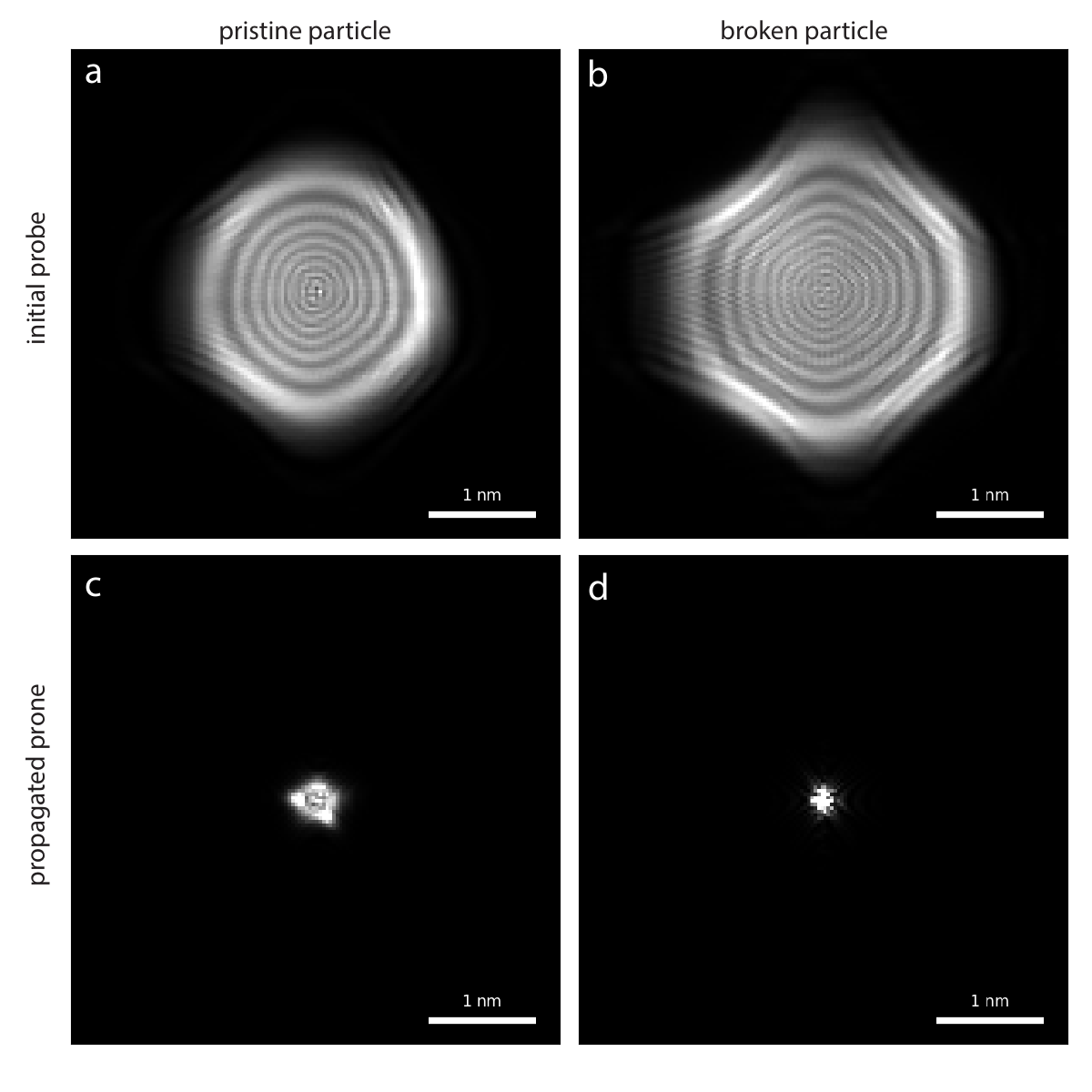}
\caption{ Reconstructed (a-b)initial and (c-d) propagated real space probes from Fig.~\ref{fig:not_broken} and Fig.~\ref{fig:broken} respectively. }
\label{fig:probes_real}
\end{figure*}

Ptychographic reconstructions were performed with the open-source py4DSTEM toolkit~\cite{varnavides2023iterative,savitzky2021py4dstem}. 
A gradient descent multislice approach with a batch update of size 110 measurements was used with 36 slices of 1.2 nm. 
The initial guess for the probe and calibration parameters came from parallax reconstructions followed by Bayesian Optimization refinement. 
Using a vacuum probe from the same session, the amplitude of the probe was constrained in Fourier space and the phase was smoothed by fitting a low-order aberration surface expansion. 
Reconstructed probes are shown in Fig.~\ref{fig:probes} \& ~\ref{fig:probes_real}. 
The object was constrained with zero-order total variation denoising along the z direction and first-order in the x-y plane, and it was also limited to be a positive potential object. 
Position correction was turned on during the reconstruction to account for small scan distortions. 
Significant regularization of both the object and probe was needed for this dose-limited experiment. 
Additional information about the reconstruction formalism can be found elsewhere~\cite{varnavides2023iterative}.  

\subsection{Structural Analysis}
\begin{figure*}[h!]
\includegraphics[width=\linewidth]{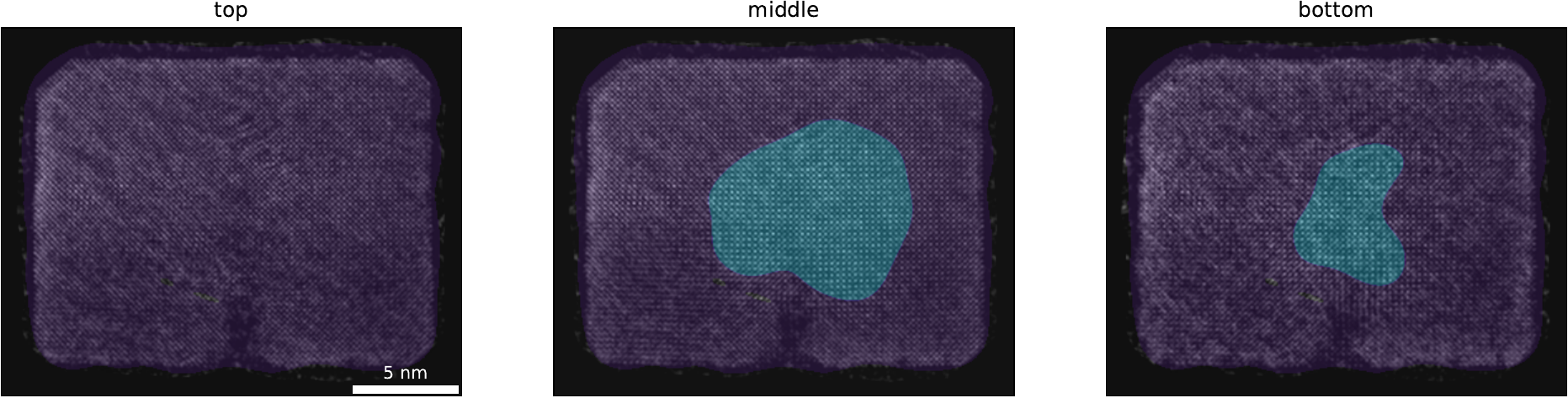}
\caption{Order parameter maps from Fig.~\ref{fig:defect}. Purple indicates shell, while blue shows core.}
\label{fig:order}
\end{figure*}

\begin{figure*}[h!]
\includegraphics[width=\linewidth]{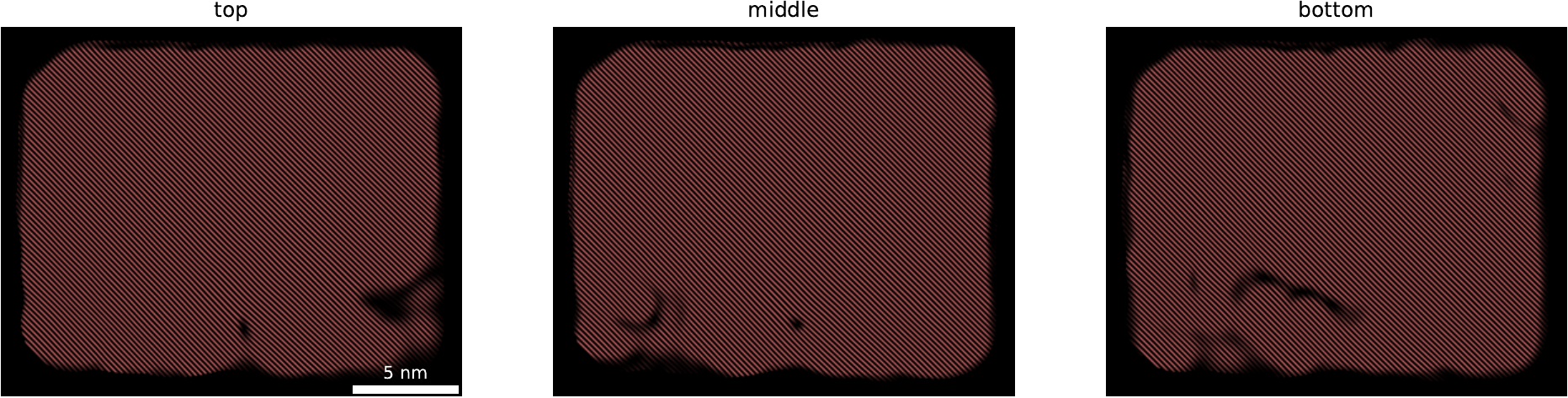}
\caption{Dislocation maps from Fig.~\ref{fig:defect}}
\label{fig:dislocation}
\end{figure*}

\begin{figure*}[h!]
\includegraphics[width=\linewidth]{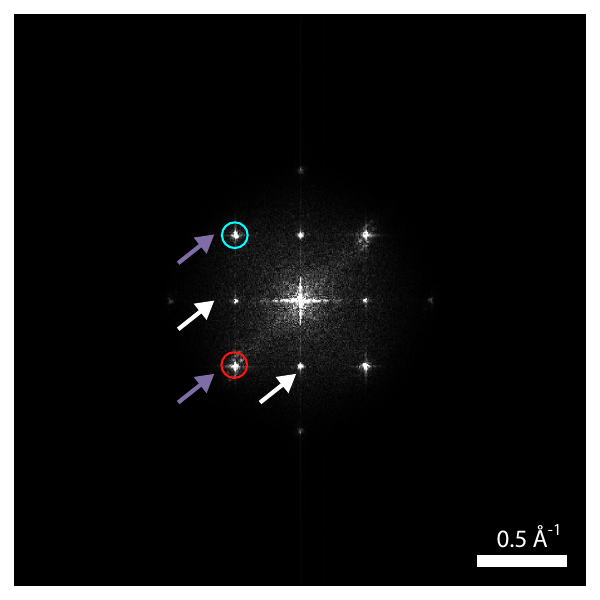}
\caption{Fourier transform of average potential in Fig.~\ref{fig:broken}. Red and blue circles indicate peaks used for dislocations plotted in Fig.~\ref{fig:defect}. The order parameter used to estimate the core was calculated from ratio of peaks indicated by white and purple arrows.}
\label{fig:fft}
\end{figure*}

Strain mapping was performed on the high-resolution lattice image from each slice. 
Fourier peak analysis was used to determine the displacement of the unit cell similar to other strain analysis methods from atomically resolved images~\cite{hytch2003measurement}. 
Strain components can be computed from the derivative of the displacement field. 
The location of the core (Fig.~\ref{fig:order}) was estimated by thresholding the ratio of the magnitude of Fourier filtered images from peaks corresponding to first and second nearest neighbors as indicated in Fig.~\ref{fig:fft} by the white and purple arrows respectively. 
In the core, the strontium and ytterbium sites are much brighter than the fluorine.
In the shell the fluorine and calcium positions are nearly identical because there are twice as many fluorine atoms as calcium atoms in each column.
This is a qualitative scoring parameter, but is nonetheless useful to get a sense of the position of the core in the volume (Fig.~\ref{fig:defect}b). 
Dislocation maps (Fig.~\ref{fig:dislocation}) were made by Fourier filtering with the peaks indicated in Fig.~\ref{fig:fft} with red and blue circles. 
The intensity of the map is from the magnitude of the Fourier peak, while the color indicates the phase, allowing visualization of breaks in the lattice. 
The 3D volume in Fig.~\ref{fig:defect}a is rendered from an intensity isosurface in x and y based on the intensity of the Fourier peaks used for strain mapping.
Only slices 8-25 are included because beyond those limits the dislocation signal becomes more difficult to track.

\begin{figure*}[h!]
\includegraphics[width=\linewidth]{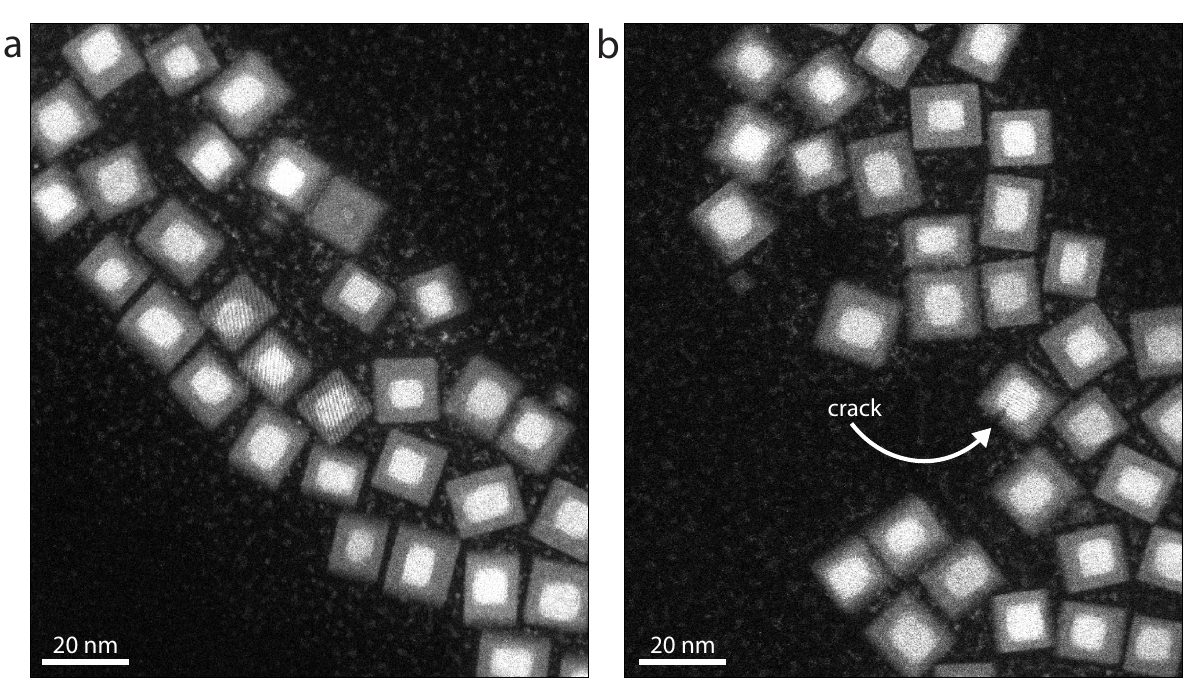}
\caption{Low magnification images of core-shell particles including one particle in (b) that is broken.}
\label{fig:lowmag}
\end{figure*}


\section{Movies}
Sample potential and 2D strain maps from all slices of particle shown in Fig.~\ref{fig:not_broken} and fig.~\ref{fig:broken}, including $\epsilon_{xx}$, $\epsilon_{yy}$, dilation ($\epsilon_{xx}$+$\epsilon_{yy}$), $\epsilon_{xy}$, and rotation. Slices 33-36 are excluded from the pristine particle movie because the intensity drops below the masking threshold. 
In both movies, the first few slices provide less reliable strain maps because of the mixed intensity from the substrate and particle.

\begin{figure*}[h!]
\includegraphics[width=\linewidth]{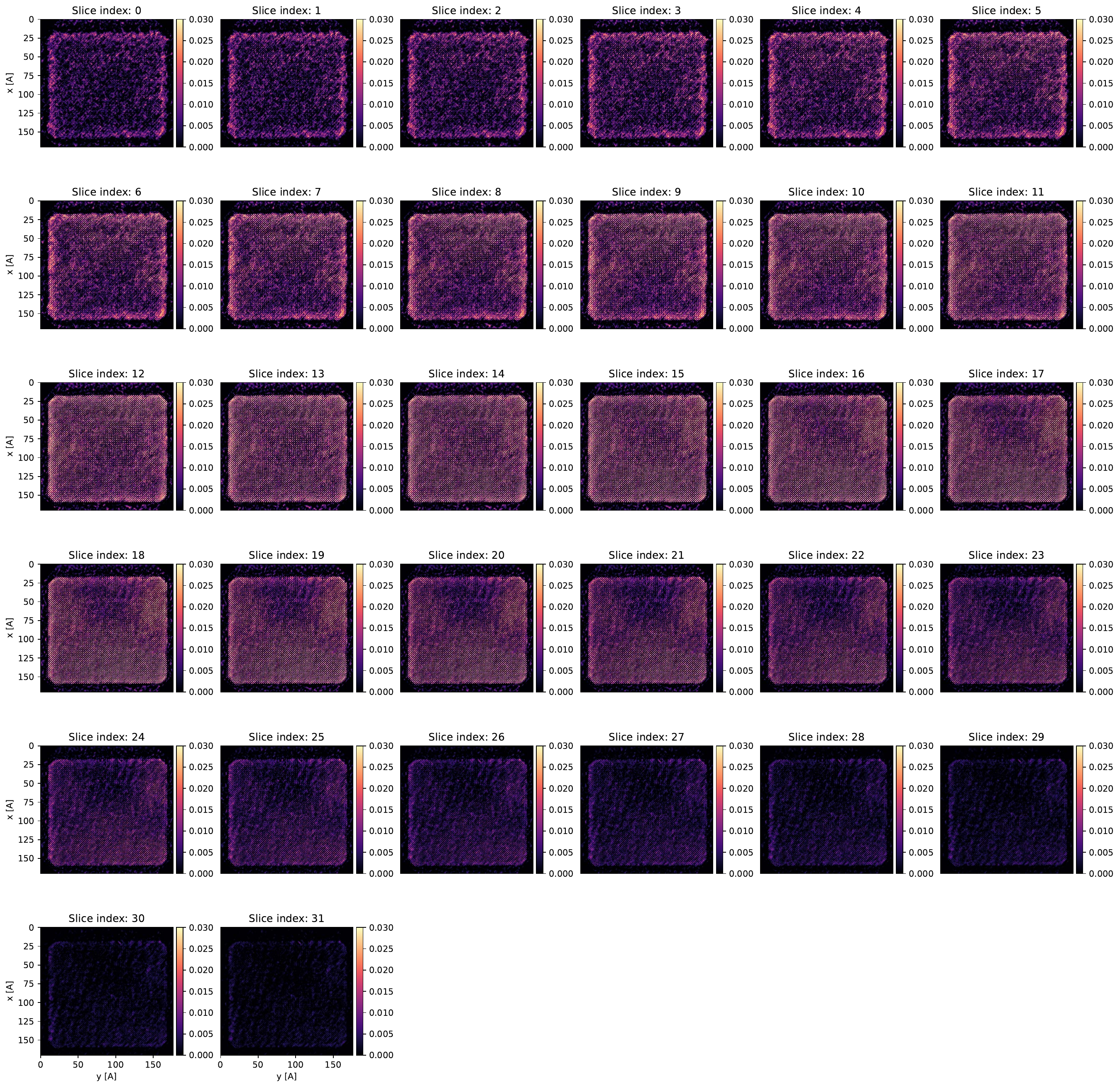}
\caption{
    All slices from ptychographic reconstruction shown in Fig.~\ref{fig:not_broken}
}
\label{fig:not_broken_slices}
\end{figure*}

\begin{figure*}[h!]
\includegraphics[width=\linewidth]{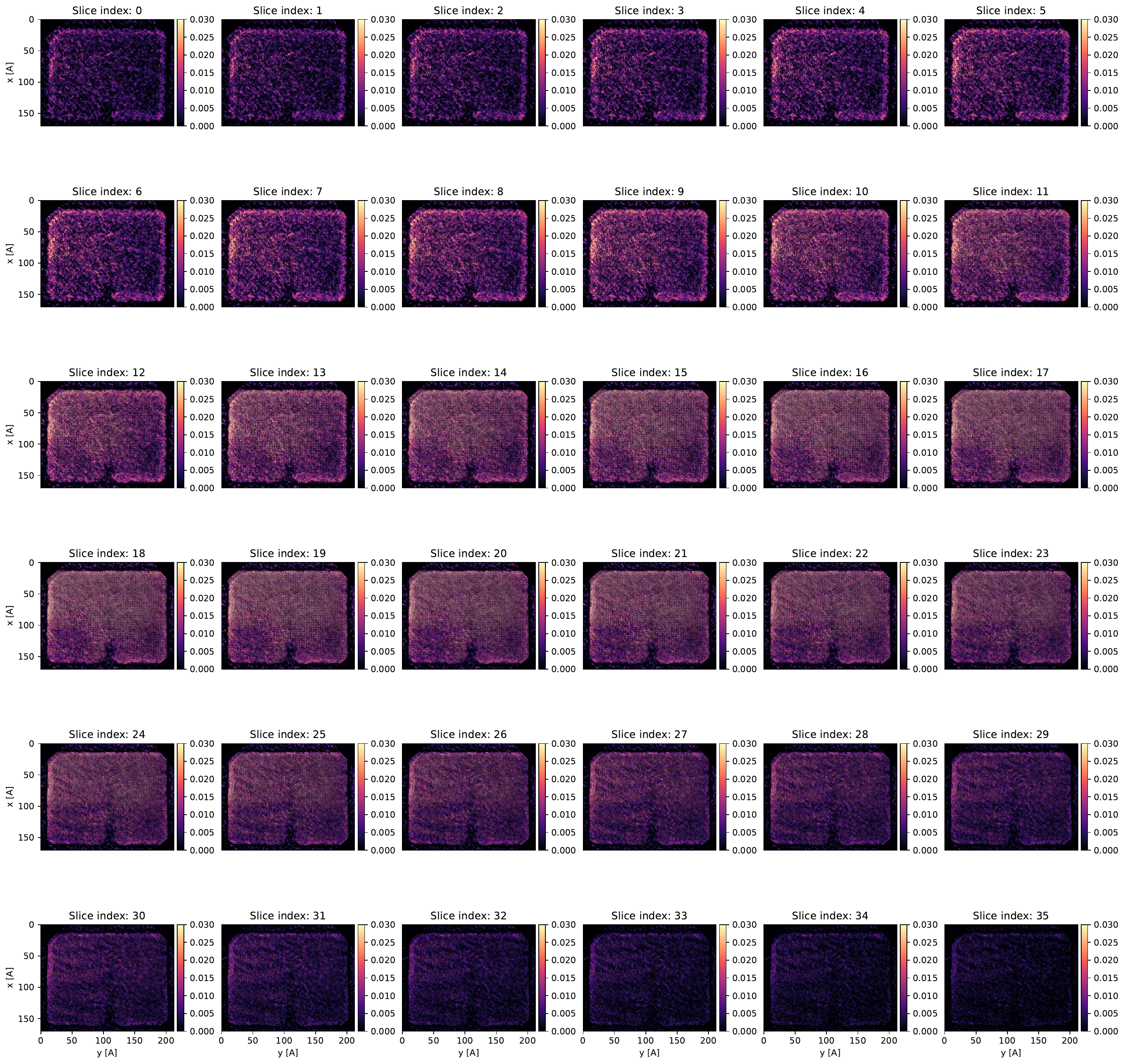}
\caption{
    All slices from ptychographic reconstruction shown in Fig.~\ref{fig:broken}
}
\label{fig:broken_slices}
\end{figure*}

\pagebreak

\end{document}